\documentclass[twoside,reqno]{HERON}
\usepackage{graphicx}
\usepackage{url}
\usepackage{amssymb,amsmath,amscd} 
\usepackage{times}
\usepackage{makeidx}
\begin{document}

\title{Nuclear symmetry energy and the PREX-CREX neutron skin puzzle within the KIDS framework}

\runningheads{Symmetry energy and neutron skin}{P. Papakonstantinou}

\begin{start}

\author{P. Papakonstantinou}{1} 

\index{Papakonstantinou, P.}

\address{Institute for Basic Science, Rare Isotope Science Project, 34000 Daejeon, Korea}{1}

\end{start} 
\begin{Abstract} 
I briefly review the KIDS theoretical framework for the nuclear equation of state (EoS) and energy density functional (EDF), I discuss recent results for the curvature parameter of the symmetry energy, and I address the PREX-CREX puzzle. 
I show that it is possible to obtain EDF models which can reproduce both PREX-II and CREX results each within its respective error bars. 
Such EDFs correspond to EoSs which soften towards low densities, as could be attributed to clusterization. 
Before such a scenario is considered viable, the dipole polarizability should also be examined. 

\end{Abstract}

\section{Introduction} 
It has proven a useful and productive practice to relate the properties of nuclear ground states and collective excitations to the nuclear equation of state (EoS)~\cite{Roca2018,garg2018}. 
For describing nuclei and also cold matter, the EoS is efficiently represented by the energy per particle 
$\mathcal{E} (\rho, \delta )$ 
as a function of baryonic density $\rho$ and isospin asymmetry $\delta$. 
Here we will consider zero-temperature unpolarized matter.  
Connections between nuclear observables and the EoS are commonly facilitated by means of energy density functional (EDF) theory~\cite{dutra2012}, 
whereby the same phenomenological EDF model is used to calculate the properties of nuclei and of infinite matter. 
As long as the relevant degrees of freedom are nucleonic, the procedure is well grounded and its success is a matter of determining the proper EDF. 
Here I will focus on non-relativistic EDFs. 

The symmetry energy is defined either as the rate of change in the energy per particle when introducing asymmetry to symmetric nuclear matter (SNM), 
\begin{equation} 
S(\rho) = \left. \frac{1}{2} \frac{\partial^2 }{\partial \delta^2} \mathcal{E} (\rho,\delta ) \right|_{\delta=0}     ,
\end{equation} 
or alternatively as the difference between the energy of pure neutron matter (PNM) and SNM,  
\begin{equation} 
S(\rho) = \mathcal{E}(\rho, 1) - \mathcal{E}(\rho, 0) .
\end{equation}
Here, I adopt the former definition. However, for the purposes of the present work the consequences of either choice are marginal.  
The symmetry energy is a very important quantity as it determines the properties and dynamics of exotic nuclei and of neutron stars~\cite{Roca2018,review2016,Oertel2017}. 
Its value at different densities is extracted through a variety of observables. 
Low densities up to saturation can be explored through nuclear data, such as masses, density distributions, collective excitations; higher densities through heavy ion collisions and astronomical observations involving compact stars. 
{\em Ab initio} calculations are also available. 

It is customary to characterize the EoS through the Taylor expansion coefficients of the energy per particle with respect to the baryonic density around the saturation density 
$\rho_0$. 
Specifically, the energy per particle in SNM and the symmetry energy are expanded as follows:
\begin{eqnarray} 
\mathcal{E} (\rho ,0)  &=& E_0 + K_0x^2/2 + Q_0x^3/6 + R_{0} x^4/24 \ldots , \\ 
S (\rho ) &=&  J + L x + K_{\mathrm{sym}} x^2/2  + Q_{\mathrm{sym}} x^3/6  + R_{\mathrm{sym}} x^4/24 \ldots  ,
\end{eqnarray} 
where $x=(\rho - \rho_0)/3\rho_0$. 
Each microscopic EDF model corresponds to a unique set of parameters $(E_0, \rho_0,K_0,\ldots,J,L,K_{\mathrm{sym}},\ldots)$. 
For most models, the higher-order expansion coefficients are not independent. This makes sense if we consider that a specific phase of matter (here: near-saturated nucleonic matter at zero temperature) must be characterized by a single adiabatic index and therefore, if each of the above expressions is cast in the approximate form $a(\rho/\rho_0)^{\gamma}$, two coefficients determine all the rest in the Taylor expansion. The concept of the adiabatic index represented here by $\gamma$ has also been very productive in studies of the symmetry energy~\cite{LyT2022}.

Focusing on the symmetry energy, going from lower to higher orders, we define the value at saturation density $J$,  the slope $L$, the curvature $K_{\mathrm {sym}}$, the skewness $Q_{\mathrm{sym}}$, the kyrtosis $R_{\mathrm{sym}}$, and so on. 
The lowest order parameters are currently estimated to lie within 
$J=30-33$~MeV and $L=40-70$~MeV, 
although they are often adjusted in light of new data or analyses - see, {\em e.g.}, \cite{Roca2018,review2016,Oertel2017,li2021,Thi2021}. 
Higher-order parameters are practically unconstrained. 
The loose constraints that do exist are indirect in that these parameters are determined, in a model-dependent way, from the lower-order ones. 

Recent developments that call for innovative ways to constrain the density dependence of the nuclear symmetry energy include multi-messenger astronomy 
and, on the nuclear-structure side, the measurements of the weak-charge form factors of $^{208}$Pb and $^{48}$Ca by the PREX-II~\cite{PREXII} and CREX~\cite{CREX} collaborations, 
respectively. 
Independent statistical analyses of astronomical observations, of {\em ab initio} calculations for neutron matter and of nuclear data reveal the relevance of $K_{\mathrm{sym}}$ or the related droplet-model parameter 
\begin{equation} 
K_{\tau} = K_{\mathrm{sym}} - \left( 6+\frac{Q_0}{K_0} \right)  L 
\end{equation} 
\cite{RaO2019,NeC2021,XuP2022}, besides the slope parameter $L$. 

In this contribution, I will focus on the tension between the PREX-II measurement and other data, especially CREX, in Sec.~\ref{Sec:Puzzle}, where I will also present first attempts at a resolution using the KIDS framework. 
But first, in Sec.~\ref{Sec:KIDS}, I will briefly introduce KIDS and in Sec.~\ref{Sec:SymEn} I will summarize part of the previous KIDS-based studies of the density dependence of the symmetry energy. 

\section{KIDS framework and the symmetry energy\label{Sec:KIDS}}
 
The Korea-IBS-Daegu-SKKU (KIDS) theoretical framework for the nuclear EoS and EDF~\cite{kidsnm,kids_nuclei1} offers the possibility to explore the EoS parameters independently of each other and independently of assumptions about the in-medium effective mass. 
It is based on a power expansion of the nuclear matter energy in terms of the Fermi momentum - equivalently, the cubic root of the density. 
The optimal number of expansion terms for describing normal and neutron star matter was found to be four terms (three for symmetric matter). 
The EoS can easily be transposed to an EDF in the form of a Skyrme functional with an extended density dependence. 
In addition to the EoS coefficients, one needs to determine also the gradient terms of the functional, represented by the isoscalar (IS) and isovector (IV) 
coupling parameters $C_{12}$ and $D_{12}$, the IS and IV in-medium effective mass, $m_s^{\ast}=\mu_sm$ and $m_v^{\ast}=\mu_vm$, represented by the 
coupling parameters $C_{\mathrm eff}$ and $D_{\mathrm eff}$, and at least one spin-orbit coupling term $W_0$ (see, e.g., \cite{kids_nuclei1}).  
A crucial novelty is that the above EDF parameters are determined without altering the EoS: any density dependence introduced to the EDF through the effective mass choice,  
is compensated for in the EoS by an equal and opposite density-dependent term of the same form $\rho^{5/3}$, while the gradient and spin-orbit terms leave no contribution to the EoS of infinite matter. 
The purpose is to be able to test any given EoS in nuclei as it is. 

Mainly three ways have been used to determine the above EDF parameters for a given EoS: 
\begin{itemize}
\item  
First, most rudimentary option: Split the EoS term $c_2\rho^{5/3}$ into a term $kc_2\rho^{5/3}$, 
which will provide the parameters $t_1,t_2$ of the Skyrme-type functional (for $x_1=x_2=0$), 
and the rest, $(1-k)c_2\rho^{5/3}$, which will provide a genuine density-dependent term.  
The optimal values of the constant $k$ and at the same time $W_0$ are determined by a fit to a minimal amount of data 
(masses and radii of three nuclei). This simple procedure typically leads to $\mu_s$ close to one. 
It is good enough for inspecting bulk nuclear and neutron-star properties~\cite{kids_nuclei1,npsm71,kids_nuclei2}, 
but is quite restrictive when looking at, {\em e.g.}, single-particle spectra and collective excitations. 
\item 
Second option: Besides the EoS, select also the desired values for the effective masses. 
Then, determine $C_{12},D_{12},W_0$ by a fit to nuclei. 
This method has been used, {\em e.g.}, in the proof-of-concept study \cite{kids_nuclei1} and in \cite{prc103}. 
\item 
Third option: Besides the EoS and the effective masses, also choose $C_{12}$, $D_{12}$, $W_0$ from the beginning. 
This is useful for inspecting trends~\cite{hnps_skin} or to reduce the parameter space~\cite{XuP2022} and I will adopt it here as well. 
\end{itemize} 
The expressions which provide the transformations from the EoS parameters to the EDF parameters can be found in the appendix of Ref.~\cite{XuP2022}.

The KIDS framework has been recently employed in efforts to constrain $K_{\mathrm{sym}}$ and $K_{\tau}$  
from both nuclear data and astronomical observations~\cite{XuP2022,prc103,GPH2022}. 
Next, I will briefly review the above efforts before turning to the PREX-CREX puzzle. 

\section{Constraints on the curvature parameter\label{Sec:SymEn}} 

A somewhat heuristic exploration of the curvature of the symmetry energy was undertaken in Ref.~\cite{prc103}. 
A standard EoS for SNM was assumed. Specifically $(\rho_0,E_0,K_0,Q_0)=(0.16$~fm$^{-3},-16$~MeV$,240$~MeV$,-373$~MeV$)$. 
Several points on the hyperplane of $(J,L,K_{\mathrm{sym}})$ were then examined with $Q_{\mathrm{sym}}$ fixed to $650$~MeV. 
For each point, the corresponding KIDS functional parameters and a pairing parameter were obtained for applications in spherical even-even nuclei. 
The different EoSs were thus tested successively on the properties of closed-shell nuclei, along the Sn isotopic chains, and on astronomical observations, in a step-by-step procedure 
of elimination and correction. 
A small regime of best-performing parameters was determined suggesting that 
that $K_{\mathrm{sym}}$ is negative and no lower than $-200$MeV 
(a subsequent analysis including $K_0$ variations suggested rather $K_{\mathrm{sym}}>-150$~MeV~\cite{GPH2022}), 
that $K_{\tau}$ likely lies between $-400$ and $-300$~MeV and that $L$ likely lies between $40$ and $65$ MeV. 
The selected EoSs show an inflection point in the symmetry energy towards two times $\rho_0$, 
a stiffening which is found necessary for obtaining realistic neutron star properties. 

The correlations among parameters or between parameters and observables were discussed in Refs.~\cite{prc103,GPH2022,hnps_skin}. 
Both $L$ and $-K_{\tau}$ were found to correlate moderately, not strongly, with the neutron skin thickness. 
$K_{\mathrm{sym}}$ was found to correlate strongly with the radius of a neutron star (heavy or canonical), while for $L$ the correlation was found moderate.  
The strong correlation observed between $K_{\mathrm{sym}}$ and $3J-L$ in the case of standard functionals~\cite{Mondal2017} is attributable to the parameter deficit of those models 
and its use for constraining $K_{\mathrm{sym}}$ from $J$ and $L$ seems misguided. 
Finally, the neutron skin thickness and the neutron star radius were found practically uncorrelated, suggesting that the former cannot be used as a predictor of the latter. 

In Ref.~\cite{XuP2022} a Bayesian analysis was performed based on a variety of observables on the nuclei $^{208}$Pb and $^{120}$Sn. 
As IV constraints, the neutron skin thickness, the electric dipole polarizability and the energy of the giant dipole resonance were used. 
IS constraints were included. 
The results for the KIDS model with up to four expansion coefficients were compared with those from the standard Skyrme functional form.  
It was found that $K_{\mathrm[sym]}$ cannot be constrained from nuclear data alone. 
It was also confirmed that for both $K_{\mathrm{sym}}$ and $K_0,Q_0$, the standard Skyrme models 
with their exceedingly tight correlations among parameters are quite constricting. 
It is then worth considering if extended models like KIDS can resolve puzzles such as the thick neutron skin of $^{208}$Pb.

\section{The PREX-CREX puzzle \label{Sec:Puzzle}} 

The neutron skin thickness $\Delta r_{np}$, defined as the difference between the root mean square radii of neutron and proton density distributions in nuclei, has been found in several studies to correlate positively with the slope $L$ of the nuclear symmetry energy or, almost equivalently, the value of the symmetry energy at around $2\rho_0/3$.  
Parity-violating electron scattering experiments have revealed a thin neutron skin in the $^{48}$Ca nucleus (CREX experiment~\cite{CREX}) and a thick neutron skin in $^{208}$Pb (PREX-II experiment~\cite{PREXII}). 
Within standard energy density functional (EDF) theory,  the two results appear irreconcilable because description of $^{48}$Ca appears to require a soft symmetry energy (low slope parameter $L$) and description of  $^{208}$Pb requires a stiff symmetry energy (high $L$)~\cite{RRN20XX,YuP20XX}. 
This remains true when uncertainties in the extraction of the neutron skin thickness from the measured parity violating asymmetry are taken into account. 

\begin{figure}[t]
\centering
\includegraphics[scale=0.31]{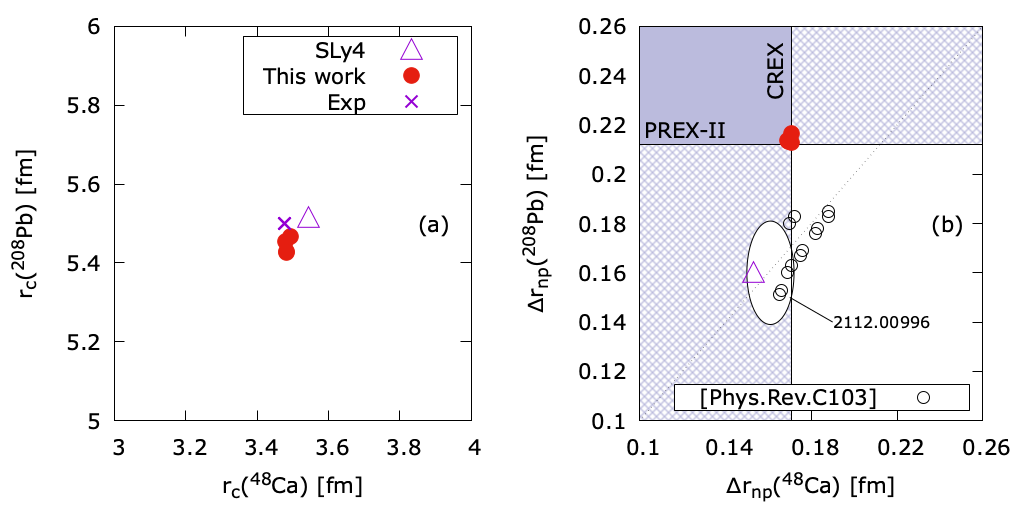}
\includegraphics[scale=0.31]{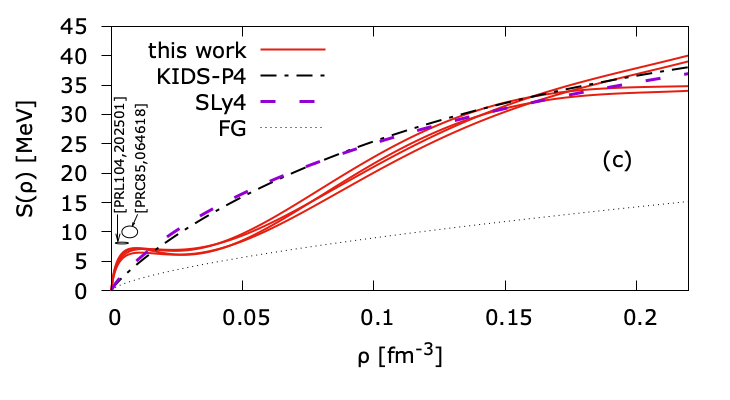}
  \caption[]{Results for the four EoS/EDF models: 
  (a) Charge radii of $^{48}$Ca and $^{208}$Pb compared with the SLy4 EDF and with measured values. 
  (b) Neutron skin thickness compared with predictions reported in \cite{prc103,skin_HCH}. 
  The shaded areas correspond to the domains of CREX ($\Delta r_{np}=0.121\pm 0.5$~fm)  and PREX-II ($\Delta r_{np}=0.283\pm 0.071$~fm). 
  (c) Corresponding density dependence of the symmetry energy compared with the KIDS-P4 model  and SLy4 as well as a free Fermi gas. 
  Values extracted experimentally for dilute matter at low temperatures~\cite{Nat2010,Wada2012} are indicated roughly with ellipses. 
  }\label{figure}
\end{figure}

As regards the standard KIDS framework, predictions for the neutron skin thickness with parameters constrained to bulk nuclear data (masses, charge radii) and neutron-star properties~\cite{prc103,GPH2022,skin_HCH} agree with the CREX measurement, but underestimate the PREX-II measurement. (See also Fig.~\ref{figure}(b).)
Meanwhile, the role of the curvature parameter has been revealed in Bayesian inference analyses, leaving open the possibility of reconciliation by a properly constrained extended EDF. 
Indeed, the analysis in \cite{XuP2022} of the PREX-II data, preliminary CREX data and other IV nuclear data within an extended energy density functional shows that the relevant posterior probability distributions for the observables of the two nuclei do show overlaps, even though the distributions are skewed to opposite directions~\cite{XuP2022}. 
However, one has yet to show that a single EDF model can reproduce the properties of both nuclei at the same time. 
Simultaneous description has been shown possible only if one allows for large confidence levels: 
In Ref.~\cite{ZhC202X}, Skyrme functionals were developed which can describe the CREX and PREX-II data to $90\%$ c.l. 

In this work, I take the CREX and PREX-II measurements at face value and ask what it takes for a nuclear EDF model to reproduce both at the same time - each within its reported error bars. 
First, I explore the EoS parameter space with the basic parameters varied within fiducial ranges, namely  
$\rho_0=0.155-0.161$~fm$^{-3}$, $-E_0=15.5-16.1$~MeV, $K_0=190-250$~MeV, 
$J=30-34$~MeV, $L=40-70$~MeV, $-K_{\tau}=250-500$~MeV, 
in $4-10$ equidistant steps each, 
while I allow the skewness parameter to vary very widely (thousands of MeVs) rather than fixing it. 
As in Ref.~\cite{XuP2022}, I fix $\mu_s$ to $0.82$ times the bare nucleon mass, a value compatible with the energy of the giant quadrupole resonance, and the enhancement factor 
$\kappa=1/\mu_v-1$
 to 0.22, compatible with a variety of IV nuclear properties~\cite{XuP2022}. 
The IS and IV gradient parameters are, respectively, varied as 
$-C_{12}=64,66,68,70$~MeV~fm$^5$ 
and fixed at 
$D_{12}=2.5$~MeV~fm$^5$. 
The spin orbit coupling strength is varied between 129 and 135~MeV~fm$^5$. 
I transpose all these parameters into a KIDS EDF model for Hartree-Fock calculations of ground-state nuclear properties.  
I find no parameter sets which can reproduce the CREX and PREX-II measurements within their error bars. 

Next, I extend the formalism to five EoS parameters, so that I can vary widely all of $Q_0, R_0, R_{\mathrm{sym}}$ as well as the lower-order ones. 
For the purposes of the present crude exploration I use 
$\rho_0=0.16$~fm$^{-3}$, 
$E_0=-16$~MeV, 
$K_0=220,240$~MeV, 
$J=32,33$~MeV, 
$L=40-70$~MeV,  
$C_{12}=-66$~MeV~fm$^5$, 
$D_{12}=2.5$~MeV~fm$^5$, 
$\mu_s=0.82$, $\kappa=0.22$, 
and $W_0=133$~MeV~fm$^5$. 
$Q_0$ (negative) and $Q_{\mathrm{sym}}$ (between $-Q_0-1000$~MeV and $-Q_0+1000$~MeV) is varied in steps of 500~MeV. 
$R_0$ and $R_{\mathrm{sym}}$ were varied in steps of $2$~GeV. 
The search should certainly be refined, but even so, I find four EoS/EDF models satisfying the following criteria: 
1) Both CREX and PREX-II results are reproduced within their reported error bars and 2) for 19 basic data on closed-shell nuclei, namely the binding energies of 
$^{16}$O, $^{40,48}$Ca, $^{56,68,78}$Ni, $^{90}$Zr, $^{100,120,132}$Sn, $^{208}$Pb, $^{218}$U 
and he charge radii of  
 $^{16}$O, $^{40,48}$Ca, $^{90}$Zr, $^{120,132}$Sn, $^{208}$Pb, the average deviation per datum (see, {\em e.g.}, \cite{prc103} for a definition) does not exceed $1\%$. 
 Results for the charge root mean square radii $r_c$ of $^{48}$Ca and $^{208}$Pb are shown in Fig.~\ref{figure}(a) and compared with data and, to demonstrate competitiveness, 
 with the standard functional SLy4~\cite{Chabanat1998}. 
 Results for the neutron skin thickness of the two nuclei are shown in Fig.~\ref{figure}(b) and compared with the KIDS predictions mentioned earlier~\cite{prc103} and with SLy4. 
 The tension is obvious, as the four new results are barely inside the CREX-PREX-II regime. 
 The symmetry energy corresponding to each of these four results is shown in Fig.~\ref{figure}(c) along with SLy4 and a previously developed KIDS parameter set, KIDS-P4, which was based on the Akmal-Pandharipande-Ravenhall~\cite{apr,kids_nuclei2} EoS. 

What the EoSs satisfying the above criteria have in common are very large skewness (order of GeV) and kyrtosis (tens of GeV). 
One can of course dismiss such a result as unnatural. 
The problem it is meant to solve, namely the tension between CREX and PREX-II, can credibly be considered spurious: 
Why should we try to reconcile the CREX and PREX-II values strictly within their (large) error bars - effectively within $1\sigma$ each? 
It is far from $100\%$ certain that the physical values lie therein. 
On the other hand, it is worth examining what the result implies. 
As shown in Fig~\ref{figure}, the EoSs correspond to a strong change in softness at low densities 
and therefore a departure from the adiabatic index of near-saturated matter: 
Instead of only one inflection point at high densities, separating the normal and the dense regimes, there appears a second one separating the normal and the dilute regimes. 
Physically, this makes sense: below $0.1$~fm$^{-3}$, nucleonic matter  becomes clusterized. 
Although the microscopic degrees of freedom are still the same nucleonic ones like at densities closer to saturation, the {\em  average} effective, in-medium  interaction among nucleons must have a different functional form in this different, clusterized medium - reflected in a different EDF and the inflected EoS.    
In this light, the result is not exotic:  
The EoSs plotted in Fig.~\ref{figure}(c) are consistent with the symmetry energy in dilute matter at low temperature as extracted by isoscaling analysis of heavy-ion collisions~\cite{Nat2010,Wada2012}. 
It is also consistent with a theoretical study of Overhauser orbitals, corresponding to $\alpha-$like clusters on a lattice~\cite{Jaq1988}. 
Then we arrive at the following prospect: The need for an interpolation of different functionals (EoSs of different softness) for dilute, normal, and dense matter as befits 
the different phases of each. 

Before this scenario can be considered viable for modeling dilute matter in nuclei, it will be important to also examine whether such highly skewed EoSs give reasonable results for the electric dipole polarizability and generally other IV observables of various nuclei. 

\section{Summary} 

The versatile KIDS theoretical framework for the nuclear EoS and EDF offers the possibility to explore the symmetry-energy parameters of low and high order independently of each other and independently of assumptions about the in-medium effective mass. 
Here, I briefly introduced the framework, I discussed recent results for $K_{\mathrm{sym}}$ and $K_{\tau}$, and I addressed the PREX-CREX puzzle. 
I showed that it is possible to obtain EDF models which can reproduce both CREX and PREX-II results each within its respective error bars. 
Such EDFs correspond to EoSs which soften towards low densities, as could be attributed to clusterization. 
The results suggest a decoupling between the density dependence at low and normal densities, similarly to previous results suggesting a decoupling of the normal and dense regimes. 
Before such a scenario is considered viable, the dipole polarizability and other IV observables should also be examined. 

\section*{Acknowledgments} 
Work supported by the Rare Isotope Science Project of the Institute
for Basic Science funded by the Ministry of Science, ICT and
Future Planning and the National Research Foundation (NRF)
of Korea (2013M7A1A1075764).



\end{document}